\documentclass[12pt]{article}
\usepackage{amsmath,amssymb,epsfig}
\makeatletter \@addtoreset{equation}{section} \makeatother
\renewcommand{\theequation}{\thesection.\arabic{equation}}
\newcommand{\ba}{\begin{array}}
\newcommand{\ea}{\end{array}}
\newcommand{\beq}{\begin{equation}}
\newcommand{\eeq}{\end{equation}}
\newcommand{\bea}{\begin{eqnarray}}
\newcommand{\eea}{\end{eqnarray}}

\def\bce{\begin{center}}
\def\ece{\end{center}}

\def\nonu{\nonumber}

\def\pa{\partial}
\def\al{\alpha}
\def\be{\beta}

\def\de{\delta}

\def\ep{\epsilon}

\def\la{\lambda}

\newcommand{\tW}{\widetilde{W}}

\newcommand{\dsqrt}[1]{\sqrt{\mathstrut #1}}

\def\eps6{{\displaystyle \mathop{\epsilon}^{6}}{}}

\def\nab6{{\displaystyle \mathop{\nabla}^{6}}{}}

\def\0{{\sst{(0)}}}
\def\1{{\sst{(1)}}}
\def\2{{\sst{(2)}}}
\def\3{{\sst{(3)}}}
\def\4{{\sst{(4)}}}
\def\5{{\sst{(5)}}}
\def\6{{\sst{(6)}}}
\def\7{{\sst{(7)}}}
\def\8{{\sst{(8)}}}

\def\ba{\begin{array}}
\def\ea{\end{array}}
\def\beq{\begin{equation}}
\def\eeq{\end{equation}}
\def\be{\begin{equation}}
\def\ee{\end{equation}}

\def\la{\lambda}
\def\eps{\epsilon}

\def\ba{\begin{array}}
\def\ea{\end{array}}
\def\beq{\begin{equation}}
\def\eeq{\end{equation}}
\def\be{\begin{equation}}
\def\ee{\end{equation}}

\def\la{\lambda}
\def\eps{\epsilon}

\def\eps6{{\displaystyle \mathop{\epsilon}^{6}}{}}

\def\nab6{{\displaystyle \mathop{\nabla}^{6}}{}}

\newcommand{\HGF}[4]{{}_2 F_1 \left(#1,#2;#3;#4\right)}

\newcommand{\bean}{\begin{eqnarray*}}
\newcommand{\eean}{\end{eqnarray*}}

\begin{document}
\thispagestyle{empty} \addtocounter{page}{-1}
   \begin{flushright}
\end{flushright}

\vspace*{1.3cm}
  
\centerline{ \Large \bf  Perturbing Around }
\vspace{.3cm} 
\centerline{ \Large \bf   A Warped Product Of
  $AdS_4$ and Seven-Ellipsoid   } 
\vspace*{1.5cm}
\centerline{{\bf Changhyun Ahn  {\rm and} Kyungsung Woo }
} 
\vspace*{1.0cm} 
\centerline{\it  
Department of Physics, Kyungpook National University, Taegu
702-701, Korea} 
\vspace*{0.8cm} 
\centerline{\tt ahn@knu.ac.kr
} 
\vskip2cm

\centerline{\bf Abstract}
\vspace*{0.5cm}

We compute the spin-2 Kaluza-Klein modes around 
a warped product of $AdS_4$ and a seven-ellipsoid. 
This background with global $G_2$ symmetry is related to a $U(N) \times
U(N)$ ${\cal N}=1$ superconformal Chern-Simons matter theory with sixth
order superpotential. The mass-squared in $AdS_4$ is quadratic in $G_2$
quantum number and KK excitation number. 
We determine the dimensions of spin-2 operators using the AdS/CFT correspondence.
The connection to ${\cal N}=2$ theory preserving $SU(3) \times U(1)_R$
is also discussed.

\baselineskip=18pt
\newpage
\renewcommand{\theequation}
{\arabic{section}\mbox{.}\arabic{equation}}

\section{Introduction}

The
three-dimensional ${\cal N}=6$ $U(N) \times U(N)$ 
Chern-Simons matter theories
with level $k$ 
can be described as the low energy limit of $N$ M2-branes at 
${\bf C}^4/{\bf Z}_k$ singularity \cite{ABJM}.
For $k=1, 2$, the full ${\cal N}=8$ supersymmetry
is preserved while for $k > 2$,
the supersymmetry is broken to ${\cal N}=6$. 
The RG flow
between the UV fixed point and 
the IR fixed point of the three-dimensional 
field theory can be found from gauged ${\cal N}=8$ 
supergravity in four-dimensions via AdS/CFT 
correspondence \cite{Maldacena}. 
The holographic
RG flow equations connecting ${\cal N}=8$ $SO(8)$ fixed point 
to ${\cal N}=2$ $SU(3) \times U(1)$ fixed point have been found in 
\cite{AP,AW} while
those from ${\cal N}=8$ $SO(8)$ fixed point 
to
${\cal N}=1$ $G_2$
fixed point also have been studied in 
\cite{AW,AI,AR99}.
The $M$-theory lifts of these 
RG flows have been constructed in \cite{CPW,AI}.

The mass deformed $U(2) \times U(2)$
Chern-Simons matter theory with level $k=1$ 
or $k=2$ preserving global $SU(3) \times U(1)_R$ symmetry has been studied 
in \cite{Ahn0806n2,BKKS,KKM,KPR} while
the mass deformation for this theory preserving $G_2$
symmetry  has been described in \cite{Ahn0806n1}. 
The nonsupersymmetric 
RG flow equations preserving $SO(7)^{\pm}$ symmetry 
have been discussed in \cite{Ahn0812}.  
The holographic
RG flow equations connecting ${\cal N}=1$ $G_2$ fixed point 
to ${\cal N}=2$ $SU(3) \times U(1)_R$ fixed point have been 
found in \cite{BHPW}.  
Moreover, the ${\cal N}=4$ and ${\cal N}=8$ RG flows have been 
studied in \cite{AW09}. Recently, further developments on  
the gauged ${\cal N}=8$ supergravity in four-dimensions have been done
in \cite{AW09-1,Ahn0905}. 

In order to understand the above 
${\cal N}=1$ mass-deformed Chern-Simons matter 
theory preserving $G_2$ symmetry(no $R$-symmetry and no chiral ring) 
in three-dimensions fully, 
the gravity dual should be used to study this strongly coupled field theory
and this is the main feature of AdS/CFT correspondence \cite{Maldacena}.
Heidenreich \cite{Heidenreich} found the complete list of $OSp(1|4)$ 
unitary irreducible representations: the structure of ${\cal N}=1$ 
supermultiplets. 
The even subalgebra of $OSp(1|4)$ is the isometry algebra of $AdS_4$.
In his classification, there exist massive supermultiplets.
Then the ${\cal N}=1$ supermultiplets of gravity states of IR theory 
at the first few KK levels can be classified according to the mass spectrum. 
The 4-dimensional KK modes are massive $AdS_4$ scalars and the masses
are determined by the eigenvalues of the differential operator acting
on 7-dimensional ellipsoid. 
Then it is necessary to compute this eigenvalue equation, i.e., both 
the eigenfunctions and the eigenvalues for given above 7-dimensional Laplacian.

In this paper, 
we compute the explicit KK spectrum of the spin-2 fields in 
$AdS_4$ by following the recent work of \cite{KPR}.
In an 11-dimensional theory, the equations for the metric
perturbations leads to a minimally coupled scalar equation. 
We obtain all the KK modes that are polynomials in the eight variables
parametrizing the deformed ${\bf R}^8$.
The squared-mass terms in $AdS_4$ for all the modes are quadratic of
the $G_2$ quantum number and the KK excitation number.
We describe the corresponding ${\cal N}=1$ dual SCFT operator
depending on the two quantum numbers.

In section 2, we review the 11-dimensional background discovered in \cite{AI}. 
In section 3, we solve the minimally coupled scalar equation in this
background and its spectrum is determined.
We match the quantum numbers of the operators 
with those of the operators in Chern-Simons matter theory with sixth
order superpotential.
In section 5, we summarize the main results of this paper and make
some comments on the future directions. 

\section{An ${\cal N}=1$ supersymmetric $G_2$-invariant 
flow in M-theory:Review}

Let us review the 11-dimensional uplift of the supergravity background 
with global $G_2$ symmetry found in \cite{Warner83} as a nontrivial extremum
of the gauged ${\cal N}=8$ supergravity in 4-dimensions.  
Our notation is as follows:the 11-dimensional coordinates
with indices $M, N, \cdots$ are decomposed into 4-dimensional spacetime 
coordinates $x^{\mu}$ with indices $\mu, \nu, \cdots$ and
7-dimensional internal space coordinates $y^m$ with indices $m, n, \cdots$.
Denoting the 11-dimensional metric as $g_{MN}$ with 
the convention 
$(-, +, \cdots, +)$
and the antisymmetric 
tensor fields as $F_{MNPQ} = 4\,\pa_{[M} A_{NPQ]}$, the bosonic
Einstein-Maxwell field equations are characterized by \cite{CJS}
\bea
R_{M}^{\;\;\;N} & = & \frac{1}{3} \,F_{MPQR} F^{NPQR}
-\frac{1}{36} \de^{N}_{M} \,F_{PQRS} F^{PQRS},
\nonu \\
\nabla_M F^{MNPQ} & = & -\frac{1}{576} \,E \,\ep^{NPQRSTUVWXY}
F_{RSTU} F_{VWXY},
\label{fieldequations}
\eea
where the covariant derivative $\nabla_M$ 
on $F^{MNPQ}$ in the second equation of 
(\ref{fieldequations})
is given by 
$E^{-1} \pa_M ( E F^{MNPQ} )$ together with elfbein determinant 
$E \equiv \sqrt{-g_{11}}$. The 11-dimensional epsilon tensor 
 $\ep_{NPQRSTUVWXY}$ with lower indices is purely numerical.
The 11-dimensional geometry is a warped product of $AdS_4$ and the
7-dimensional ellipsoid.
We refer the reader to \cite{AI,AI02} for a derivation of the formula
in this section.

The gauged ${\cal N}=8$ supergravity theory has self-interaction of a single
massless ${\cal N}=8$ supermultiplet of spins 
$(2, \frac{3}{2}, 1, \frac{1}{2}, 0^{+}, 0^{-})$ with local $SO(8)$ and  local
$SU(8)$ invariance. 
There exists a non-trivial effective potential for the scalars
that is  proportional to the square of the $SO(8)$ gauge coupling $g$.  
The 70 real, physical 
scalars characterized by $(0^{+}, 0^{-})$ of ${\cal N}=8$ supergravity
parametrize the coset space $E_{7(7)}/SU(8)$ and they are described by 
an element ${\cal V}(x)$ of the fundamental 56-dimensional representation
of $E_{7(7)}$. 
Any ground state leaving the symmetry unbroken is necessarily
$AdS_4$ space with a cosmological constant proportional to $g^2$. 
Turning on the
scalar fields proportional to the self-dual tensor $C_{+}^{IJKL}$ of
$SO(8)$ yields an 
$SO(7)^{+}$-invariant vacuum while 
turning on pseudo-scalar fields proportional to
the anti-self-dual tensor $C_{-}^{IJKL}$ of $SO(8)$ 
yields $SO(7)^{-}$-invariant vacuum. Both $SO(7)^{\pm}$
vacua are nonsupersymmetric. However, simultaneously turning on
both scalar and pseudo-scalar fields proportional to 
$C_{+}^{IJKL}$ and $C_{-}^{IJKL}$, respectively, one obtains $G_2$-invariant
vacuum with ${\cal N}=1$ supersymmetry \cite{dNW}. 
The most general vev of 56-bein preserving $G_2$-invariance can be parametrized by
\bea
\phi_{IJKL} = \frac{\la}{2\sqrt{2}} \left( \cos \al \;\; C_{+}^{IJKL}+
i \sin \al \;\; C_{-}^{IJKL} \right). 
\label{phiijkl}
\eea

The two vevs $(\lambda, \alpha)$ in (\ref{phiijkl}) are given by functions 
of the $AdS_4$ radial coordinate $r \equiv x^4$. The metric formula of 
\cite{dNW} 
generates the 7-dimensional metric from the two input data of $AdS_4$ vevs 
$(\lambda,\alpha)$. The $G_2$-invariant RG flow
is a trajectory in $(\lambda,\alpha)$-plane and is parametrized by the 
$AdS_4$ radial coordinate $r$. 
Instead of using $(\lambda,\alpha)$, it is convenient to use 
$(a,b)$ defined by \cite{AI}
\begin{eqnarray}
a &\equiv& \cosh\!\left(\frac{\lambda}{\sqrt{2}}\right)
+\cos\alpha\,\sinh\!\left(\frac{\lambda}{\sqrt{2}}\right),\nonumber\\
b &\equiv& \cosh\!\left(\frac{\lambda}{\sqrt{2}}\right)
-\cos\alpha\,\sinh\!\left(\frac{\lambda}{\sqrt{2}}\right).
\label{abvevs}
\end{eqnarray}
Let us introduce the standard metric of a 7-dimensional ellipsoid. 
Using the diagonal matrix $Q_{AB}$ given by \cite{dNW,AI}
\bea
Q_{AB}={\rm diag}\left(1, 1, 1, 1, 1, 1, 1, \frac{a^2}{b^2}\right),
\label{QQ}
\eea
the 7-dimensional ellipsoidal metric with the eccentricity 
$\sqrt{1-\frac{b^2}{a^2}}$ can be written as
\begin{equation}
ds_{EL(7)}^2 = dX^A Q^{-1}_{AB}\,dX^B 
-\frac{b^2}{\xi^2}\left(X^A \delta_{AB}\,dX^B\right)^2, \label{7dmet2}
\end{equation}
where the ${\bf R}^8$ coordinates $X^A (A=1,\dots,8)$
are constrained on the unit round ${\bf S}^7$, that is, 
$\sum_{A=1}^8 (X^A)^2 =1$, 
and $\xi^2  \equiv b^2 \,X^A Q_{AB} X^B$ with (\ref{QQ}) is a quadratic form 
on the 7-dimensional ellipsoid. 
The standard metric (\ref{7dmet2}) can be rewritten, using the
explicit realization between $X^A$ and $y^m$,  
in terms of the 
7-dimensional coordinates $y^m$ such that
\begin{equation}
ds_{EL(7)}^2 = \frac{\xi^2}{a^2} \,d\theta^2 
+\sin^2 \theta\,d\Omega_6^2, \label{7dmet3}
\end{equation}
where $\theta \equiv y^7$ is the fifth coordinate in 11-dimensions 
and the quadratic form $\xi^2$ is  given by
\bea
\xi^2 =a^2 \cos^2 \theta + b^2 \sin^2 \theta,
\label{xi}
\eea
which turns to be 1 for the round ${\bf S}^7$ with $(a,b)=(1,1)$ which
has $SO(8)$ symmetry. For other values of $a$ and $b$, the $SO(8)$
symmetry group is broken down generically to $G_2$. 
The metric on the round ${\bf S}^6 \simeq \frac{G_2}{SU(3)}$ 
is denoted by $d\Omega_6^2$. 
The geometric parameters $(a,b)$ for the 7-dimensional ellipsoid can be identified 
with the two vevs $(a,b)$ defined in (\ref{abvevs}). 
This is one of the reasons why we have introduced 
$(a,b)$ in (\ref{abvevs}) rather than the original vevs $(\lambda,\alpha)$. 

Applying the Killing vector together with 
the 28-beins to the metric formula \cite{dNW}, one obtains the 
inverse metric $g^{mn}$ including the warp factor $\Delta$ not yet determined. 
Substitution of this inverse metric into the definition of warp factor $\Delta$
\cite{dNW} provides a self-consistent equation for $\Delta$. 
For the $G_2$-invariant RG flow, solving this equation yields the warp factor
\bea
\Delta=a^{-1}\,\xi^{-{4 \over 3}},
\label{delta}
\eea
where $\xi$ is given by (\ref{xi}). Then we substitute this warp factor 
into the inverse metric to obtain the 7-dimensional warped ellipsoidal
metric as follows \cite{AI,AI02}: 
\begin{equation}
ds_7^2 =g_{mn}(y) \,dy^m dy^n
=\sqrt{\Delta\,a} \, L^2 
\left(\frac{\xi^2}{a^2}\,d\theta^2 
+\sin^2 \theta\,d\Omega_6^2\right)=
\sqrt{\Delta\,a} \, L^2 \, ds_{EL(7)}^2, \label{ellip}
\end{equation}
where one can see that the standard 7-dimensional ellipsoidal 
metric (\ref{7dmet3}) 
is warped by a factor $\sqrt{\Delta\,a}$.
The nonlinear metric ansatz finally combines the 7-dimensional metric 
(\ref{ellip}) with the four dimensional metric with warp factor to yield
the 11-dimensional warped metric with (\ref{delta}), (\ref{xi}) 
and (\ref{ellip}) that solves (\ref{fieldequations}) with the
appropriate 4-form field strengths below: 
\begin{equation}
ds_{11}^2 =\Delta^{-1}\left(dr^2 +e^{2 A(r)}
\eta_{\mu\nu}dx^\mu dx^\nu \right)+ds_7^2,
\label{11dmet}
\end{equation}
where $r \equiv x^4$ and $\mu,\nu=1,2,3$ with $\eta_{\mu\nu}={\rm
  diag}(-,+,+)$. 
The $(a,b)$ are set to 
\bea
a=\sqrt{6 \sqrt{3} \over 5},\quad b=\sqrt{2 \sqrt{3} \over 5},
\label{ab}
\eea
for $G_2$-invariant IR critical point.

The 3-form gauge field with 3-dimensional M2-brane indices may be defined 
by \cite{CPW}
\begin{equation}
A_{\mu\nu\rho} = -e^{3A(r)}\,\tW(r,\theta)\,
\epsilon_{\mu\nu\rho},\label{ansatz1}
\end{equation}
where $\tW(r, \theta)$ is a geometric superpotential \cite{KW} 
to be determined. 
The $\theta$-dependence of $\tW(r, \theta)$ was essential 
to achieve the $M$-theory lift of the RG flow.
As performed in \cite{GW}, 
the $G_2$-covariant tensors living on the round ${\bf S}^6$ can be
obtained by using the 
imaginary octonion basis of ${\bf S}^6$. 
Thus we arrive at the 
most general $G_2$-invariant ansatz \cite{AI}:
\begin{eqnarray}
A_{4mn} = g(r,\theta)\,F_{mn}, \qquad
A_{5mn} = h(r,\theta)\,F_{mn}, \qquad
A_{mnp} = h_1 (r,\theta)\,T_{mnp} +h_2 (r,\theta)\,S_{mnp},\label{ansatz2}
\end{eqnarray}
where $m,n,p$ are the ${\bf S}^6$ indices and run from 6 to 11. 
The almost complex structure on the ${\bf S}^6$ is denoted by $F_{mn}$ 
which obeys $F_{mn}F^{\,nl}=-\delta_m^{\,l}$. 
The $S_{mnp}$ is the parallelizing torsion of the unit round ${\bf S}^7$ 
projected onto the ${\bf S}^6$, 
while the $T_{mnp}$ denotes the 6-dimensional Hodge dual of $S_{mnp}$.
We refer to \cite{GW} for further details about these tensors. 
The above ansatz for gauge field is the most general one which 
preserves the $G_2$-invariance and is consistent with the 11-dimensional 
metric (\ref{11dmet}).

Through the definition $F_{MNPQ}\equiv 4\,\partial_{[M}A_{NPQ]}$, 
the ansatz (\ref{ansatz1}) generates the field strengths
\begin{eqnarray}
F_{\mu\nu\rho 4} = e^{3A(r)}
\,W_r (r,\theta)\,\epsilon_{\mu\nu\rho},\qquad
F_{\mu\nu\rho 5} = e^{3A(r)}
\,W_{\theta}(r,\theta)\,\epsilon_{\mu\nu\rho},\label{fst1}
\end{eqnarray}
while the ansatz (\ref{ansatz2}) provides
\begin{eqnarray}
F_{mnpq} &=& 2\,h_2 (r,\theta)\,\eps6_{mnpqrs} F^{rs}, \qquad
F_{5mnp} = \tilde{h}_1 (r,\theta)\,T_{mnp}
+\tilde{h}_2 (r,\theta)\,S_{mnp}, \nonumber\\
F_{4mnp} &=& \tilde{h}_3 (r,\theta)\,T_{mnp}
+\tilde{h}_4 (r,\theta)\,S_{mnp}, \qquad
F_{45mn} = \tilde{h}_5 (r,\theta)\,F_{mn}, \label{fst2}
\end{eqnarray}
where the coefficient functions which depend on both $r$ and $\theta$
are given by \cite{AI}
\begin{eqnarray}
W_r &=& e^{-3A}\partial_r (e^{3A}\tW),\qquad 
W_\theta ~=~ e^{-3A}\partial_\theta (e^{3A}\tW),\nonumber\\
\tilde{h}_1 &=& \partial_\theta h_1 -3h,\qquad
\tilde{h}_2 ~=~ \partial_\theta h_2, \qquad
\tilde{h}_3 = \partial_r h_1 -3g,\nonu \\
 \tilde{h}_4 & = & \partial_r h_2, \qquad
\tilde{h}_5 = \partial_r h -\partial_\theta g. \label{fst3}
\end{eqnarray}
The mixed 
field strengths $F_{\mu\nu\rho5}$, $F_{4mnp}$ and $F_{45mn}$ were new.  
At both $SO(8)$-invariant UV and $G_2$-invariant IR critical points, 
the 4-dimensional spacetime becomes asymptotically $AdS_4$ and the mixed 
field strengths should vanish there. 
In particular, the $F_{\mu\nu\rho5}$ and $F_{45mn}$ are proportional to 
$W_\theta$ and $\tilde{h}_5$, respectively, so that they must be subject to 
the non-trivial boundary conditions:
$W_\theta =0$ and  $\tilde{h}_5=0$,
at both UV and IR critical points. 
It was checked that the $F_{4mnp}$ goes to zero 
at both critical points without imposing any boundary condition.

Applying the field strength ansatz (\ref{fst1}), (\ref{fst2}) and the metric 
(\ref{11dmet}) to the 11-dimensional Maxwell equation in 
(\ref{fieldequations}), we 
have obtained
\begin{eqnarray}
\tilde{h}_1 &=& 2L\, a^{-3} \,\xi^{-2}\,W_r h_2 
-\frac{1}{4}\,L^2  \, a^{-3}\, \xi^{-2} \, 
 e^{-3A} \pa_r (a^2 e^{3A} \sin^2 \theta
\tilde{h}_5), \nonumber\\
 \tilde{h}_3 &=& -2L^{-1}\,\xi^{-2}\,W_\theta h_2 
+\frac{1}{4}\,  \xi^{-2}  
 e^{-3A} \pa_{\theta} (a^2 e^{3A} \sin^2 \theta
\tilde{h}_5).
\label{maxf2}
\end{eqnarray}
The 11-dimensional 
Einstein-Maxwell equations were checked from the warped metric (\ref{11dmet}) 
and the field strength ansatz (\ref{fst1}) and (\ref{fst2}). 
The 11-dimensional field equations were closed within the field strengths 
$W_r$, $W_\theta$, $h_2$ and $\tilde{h}_5$ (\ref{fst3}) 
although they cannot be solved 
separately without imposing certain ansatz for them. 
Solving the ansatz one obtained 
\begin{eqnarray}
h_2 &=& \frac{L^3}{2} \dsqrt{ b^2\,(ab-1)}\,\,\xi^{-2} \sin^4 \theta, \nonumber\\ 
\tilde{h}_5 &=& 2 L^2 \dsqrt{\dfrac{(ab-1)}
{(a^2 +7b^2)^2 -112\,(ab-1)}}\,\,(-4a +a^2 b +7b^3) \,a^{-{1 \over 2}} \sin^2 \theta, 
\label{msol12}
\end{eqnarray}
and
\begin{eqnarray}
W_r &=& -\frac{1}{2L}\,a^2 \left[
a^5 \cos^2 \theta +a^2 b\,(ab-2)\,(4+3\cos 2\theta)
+b^3 \,(7ab-12)\sin^2 \theta \right], \nonumber\\
W_\theta &=& -\frac{a^{\frac{3}{2}} \,
\left[48\,(1-ab)+(a^2 -b^2)\,(a^2 +7b^2)\right]}
{
\dsqrt{(a^2 +7b^2)^2 -112\,(ab-1)}}\,\sin \theta \cos \theta. 
\label{msol34}
\end{eqnarray}
It turned out that 
the solutions (\ref{msol12}) and (\ref{msol34}) 
actually consist of an exact solution to the 11-dimensional supergravity, 
provided that the deformation parameters $(a,b)$ of the 
7-ellipsoid and the domain wall amplitude $A(r)$ 
develop in the $AdS_4$ radial 
direction along the $G_2$-invariant RG flow. 
Finally, the geometric superpotential in (\ref{ansatz1}) yields
\begin{equation}
\tW =\frac{a^{\frac{3}{2}} \,
\left(\left[48\,(1-ab)+(a^2 -b^2)\,(a^2 +7b^2)\right]\cos^2 \theta
+8\,(1 -a b) +b^2 \,(a^2 +7 b^2) \right)}{2 \dsqrt{(a^2 +7b^2)^2 -112\,(ab-1)} }.
\label{geom}
\end{equation}
Therefore, all the field strengths in (\ref{fst1}) and (\ref{fst2})
are determined via (\ref{msol12}), (\ref{msol34}) and (\ref{maxf2}). 
The 3-form gauge field (\ref{ansatz1}) 
is also determined via (\ref{geom}).

What is the dual gauge theory? Let us recall the $U(2) \times U(2)$
Chern-Simons matter theory where the matter fields consist of 
seven flavors $\Phi_i$($i=1, 2, \cdots, 7$) transforming in the
adjoint with flavor symmetry $G_2$. 
There exist standard Chern-Simons terms with levels for the gauge
groups $(k, -k)$ with $k=1,2$. 
The $\Phi_i$ forms a
septet ${\bf 7}$ of the ${\cal N}=1$ theory. When 
we turn on the mass perturbation in the gauged 
supergravity, the dual theory flows from the UV to the IR. In the dual field
theory, one integrates out the massive field $\Phi_8$(which is a
singlet ${\bf 1}$ of $G_2$ with adjoint index) characterized by
the superpotential $\frac{1}{2} m \Phi_8^2$, at a low enough
scale, and then  
this results in the sixth order superpotential \cite{Ahn0806n1}.

\section{KK spectrum of minimally coupled scalar }

Let us assume the system that a minimally coupled scalar field 
is interacting with the gravitational field. The action for 
a minimally coupled scalar field in the background which is a warped
product of $AdS_4$ and 7-dimensional ellipsoid is given by \cite{KPR}
\bea
S = \int d^{11} x \sqrt{- g} \left[ -\frac{1}{2} (\pa \phi)^2 \right].
\label{act}
\eea
The equation of motion from this action (\ref{act}) is given by
\bea
\Box \phi =0,
\label{eqm}
\eea
where $\Box$ is the 11-dimensional Laplacian. 
Using the separation of variables
\bea
\phi = \Phi(x^{\mu}, r) Y(y^m),
\label{phi}
\eea
and substituting (\ref{phi}) into (\ref{eqm}), 
one writes (\ref{eqm}) as
\bea
Y(y^{m}) \Box_4 \Phi(x^{\mu}, r) + \Phi(x^{\mu}, r) {\cal L}
Y(y^{m}) =0,
\label{sol1}
\eea
where $\Box_4$ denotes the $AdS_4$ Laplacian and 
${\cal L}$ denotes a differential operator acting on  
7-dimensional ellipsoid and is given by
\bea
{\cal L}  \equiv \frac{\Delta^{-1}}{\sqrt{-g_{11}}} \, \pa_{M} \left( 
\sqrt{-g_{11}} \, g_{11}^{M N} \, \pa_{N} \right)
= \frac{\Delta^{-\frac{3}{4}}}{\sqrt{g_{7}}} \, \pa_{m} \left(
  a^{-\frac{1}{2}} \,L^{-2} \,
\Delta^{-\frac{3}{4}} \, \sqrt{g_{7}} \, g_{7}^{m n} \, \pa_{n} \right),
\label{diffop}
\eea
where $g_{mn}^7$ and $g_{MN}^{11}$ are given by the metrics (\ref{7dmet2})
and 
(\ref{11dmet}) respectively.
In particular, the 7-dimensional metric is given by
\bea
g_{mn}^7 =
\left(
\begin{array}{ccccccc}
\frac{1}{3}(2+c_{2\theta}) & 0 &0 &0 &0 &0 &  0  \\ 
0 & s^2_{\theta} s^2_{\theta_6} &0 &0 &0 &0 &  0 \\
0 & 0 &\frac{1}{4} s^2_{\theta} s^2_{\theta_1} s^2_{\theta_6} &0 &0 &0 &  0 \\
0 & 0 &0 & \frac{1}{4} s^2_{\theta} s^2_{\theta_1} s^2_{\theta_6}
&
\frac{1}{4} c_{\alpha_1} s^2_{\theta} s^2_{\theta_1} s^2_{\theta_6} &
\frac{1}{2} c_{\alpha_1} s^2_{ \theta} s^2_{ \theta_1} s^2_{ \theta_6} &  0 \\
0 & 0 &0 & \frac{1}{4} c_{\alpha_1} s^2_{\theta} s^2_{\theta_1} s^2_{\theta_6}
&
\frac{1}{4}  s^2_{\theta} s^2_{\theta_1} s^2_{\theta_6} &
\frac{1}{2} s^2_{ \theta} s^2_{ \theta_1} s^2_{ \theta_6}
&  0 \\
0 & 0 &0 & \frac{1}{2} c_{\alpha_1} s^2_{\theta} s^2_{\theta_1} s^2_{\theta_6}
&
\frac{1}{2} s^2_{\theta} s^2_{\theta_1} s^2_{\theta_6} &
s^2_{ \theta}  s^2_{ \theta_6} &  0 \\
0 & 0 &0 &0 &0 &0 &  s^2_{\theta}, 
\end{array} \right),
\label{7dmetric}
\eea
where we use the simplified notations $c_{2\theta}\equiv \cos 2\theta$
and $s_{\theta} \equiv \sin \theta$ and so on. Let us introduce the
angular coordinates $y^m \equiv (\theta, \theta_1, \alpha_1, \alpha_2,
\alpha_3, \theta_5, \theta_6)$ \cite{AI} parametrizing the ${\bf S}^7$ 
inside ${\bf R}^8$ with $\sum_{A=1}^{8} (X^A)^2=1$ where
the relation to the rectangular coordinates is  
\bea
X^1 + i X^2  & = & \sin \theta \sin \theta_6 \sin \theta_1 \cos
(\frac{1}{2} \alpha_1) e^{\frac{i}{2}(\alpha_2 +\alpha_3)} e^{i\theta_5},
\nonu \\
X^3 + i X^4  & = & \sin \theta \sin \theta_6 \sin \theta_1 \sin
(\frac{1}{2} \alpha_1) e^{-\frac{i}{2}(\alpha_2 -\alpha_3)} e^{i\theta_5},
\nonu\\
X^5 + i X^6  & = & \sin \theta \sin \theta_6  \cos
\theta_1  e^{i\theta_5}, \nonu \\
X^7 & = & \sin \theta \cos \theta_6, \nonu \\
X^8 & = & \cos \theta.
\label{rect}
\eea
This is ${\bf R}^8$ embedding of ${\bf S}^7$ with $\frac{G_2}{SU(3)}$ base
and the relation to the Hopf fibration on ${\bf CP}^3$ is given in \cite{AI}. 
The quadratic form (\ref{xi}) is given by $\xi^2 =\frac{2}{5} \sqrt{3}
(2+ c_{2\theta})$ 
and the warp factor (\ref{delta}) is given by 
$\Delta=\frac{5^{\frac{7}{6}}}{2^{\frac{7}{6}}\times 
3^{\frac{13}{12}}(2+c_{2\theta})^{\frac{2}{3}}}$.
The equation of ellipsoid is $\frac{5\sum_{A=1}^7 (X^A)^2 }{6\sqrt{3}} + 
\frac{5(X^8)^2}{2\sqrt{3}}=1$.

Let us find out the eigenfuction $Y(y^m)$ of the differential operator 
${\cal L}$
\bea
{\cal L} Y(y^m) = -m^2 Y(y^m).
\label{diff1}
\eea
Then the equation (\ref{sol1}) leads to the equation of motion of a
massive scalar field in $AdS_4$ as follows:
\bea
 \Box_4 \Phi(x^{\mu}, r) -m^2 \Phi(x^{\mu}, r) = 0. 
\label{ads4}
\eea
Therefore, the 11-dimensional minimally coupled scalar provides a
tower of KK modes which are all massive scalars (\ref{ads4}) with
masses 
$m^2$ determined
by the eigenvalues of the above differential operator ${\cal L}$. 

Let us recall that the $G_2$ symmetry is the isometry group of the
metric.
The isometry of round ${\bf S}^6$ only is given by $SO(7)$ which
contains $G_2$ as a subgroup
and the Killing vector associated to the $G_2$ symmetry is given by
\bea
K^{a} = K_A^{~a} \pa^A
=\left[\,X^B (T^{a})_{BA} -X^C (T^{a})_{AC}\right] \pa^A, 
\label{Killing}
\eea
where $A, B = 1, 2, \cdots, 7$ for rectangular coordinates, 
$a=1, 2, \cdots, 14$ for adjoint indices, and
$T^a$ are traceless antihermitian matrices and the generators of
$G_2$.
The dimension of $G_2$ is $14$.
The explicit form of these is given by (\ref{gen}) and (\ref{gen1}) of
Appendix A.
We have checked that the metric $g_{mn}^7$ (\ref{7dmetric}) has 
vanishing Lie derivative \cite{DNP} 
with respect to the Killing vector fields $K^a$.
That is, $K^{p\,a} \pa_{p} g_{mn}^7 + (\pa_m K^{p\,a}) g_{pn}^7 + 
(\pa_n K^{p\,a}) g_{mp}^7=0$ where $K^a = 
K_q^{~a} \pa^q = g_{qp} K^{p\,a} \pa^q$. Note that in this check, we
need to work with the angular coordinates using the chain rules
between the rectangular coordinates and angular coordinates
(\ref{rect}):
after introducing the radial coordinate and obtaining the Jacobian, 
this radial coordinate is set to one.

In an irreducible representation of $G_2$ labeled by the Dynkin
labels $(\lambda, \mu)$,
the quadratic Casimir operator of $G_2$
\bea
{\cal C}_2 \equiv \sum_{a=1}^{14} (K^{a})^2, 
\label{c2}
\eea
has eigenvalues \cite{Macfarlane,Okubo}
\bea
{\cal C}_2(\lambda, \mu) = -16 
\left(\la^2 + \frac{1}{3} \mu^2 + \la \mu + 3 \la +\frac{5}{3} \mu \right),
\label{c2new}
\eea
for $(\lambda,\mu)$ representation. Note that $\sum_{a=1}^{14} (T^{a})^2 =-8$.
The explicit form for (\ref{c2}) is given by (\ref{c2appen}) of Appendix A.

The spin-2 massive ${\cal N}=8$ supermultiplet \cite{DNP} at level $n$ is characterized
by the $SO(8)$ Dynkin labels $(n,0,0,0)$, this breaks into the $SO(7)$
Dynkin labels $(0,0,n)$, and finally the massive multiplets of ${\cal
  N}=8$ for $n=1,2, \cdots$, are decomposed into $(0,0) \oplus (0,1)
\oplus \cdots
\oplus (0,n)$ under the $G_2$ symmetry. In particular, one has, with
the help of \cite{ps},  
\bea
SO(8) & \rightarrow & SO(7)   \rightarrow G_2, \nonu \\
{\bf 8}_v(1,0,0,0) & \rightarrow & {\bf 8}(0,0,1)   \rightarrow {\bf 1}(0,0) \oplus
{\bf 7}(0,1),
\nonu \\
{\bf 35}_v(2,0,0,0) &\rightarrow & {\bf 35}(0,0,2)  \rightarrow {\bf 1}(0,0)
\oplus {\bf 7}(0,1) \oplus {\bf 27}(0,2), \nonu \\
{\bf 112}_v(3,0,0,0) & \rightarrow & {\bf 112}'(0,0,3) \rightarrow {\bf
  1}(0,0)
\oplus {\bf 7}(0,1) \oplus {\bf 27}(0,2) \oplus {\bf 77}'(0,3), 
\nonu \\
(n,0,0,0) & \rightarrow & (0,0,n) \rightarrow (0,0)
\oplus (0,1) \cdots \oplus (0,n). 
\label{sog2} 
\eea
Then the relevant eigenvalues for quadratic Casimir from (\ref{c2new}) 
when $\lambda=0$ are given by 
\bea
{\cal C}_2(0, \mu) = -\frac{16}{3} \mu \left( \mu + 5 \right).
\label{eigenvalues}
\eea
Note that the factor $\mu \left( \mu + 5 \right)$ occurs in the
quadratic Casimir operator of $SO(7)$.

What are the corresponding eigenfunctions?
It is well-known that the scalar spherical harmonic for $D$-sphere 
${\bf S}^D$ is characterized
by each independent component of totally symmetric traceless tensor of
rank $n$.  Namely, they are linear combinations of products of $\mu$
factors of $X^A$'s
\bea
Y_{(0, \mu)}(X^A, A\neq 8) = C_{i_1 i_2 \cdots i_{\mu}}^{(0,\mu)} 
X^{i_1} X^{i_2} \cdots X^{i_{\mu}},
\label{sol}
\eea
where $C_{i_1 i_2 \cdots i_{\mu}}^{(0,\mu)}$ is a $(0,\mu)$-tensor
independent of the $X^A$'s and is symmetric in $i_1 i_2 \cdots
i_{\mu}$.
Furthermore, they are traceless in the sense that $C_{i_1 i_2 \cdots
  i_{\mu}}^{(0,\mu)}
\delta^{i_m i_n} =0$ for any $1\leq m, n \leq \mu$ \cite{LMRS}.  
Since ${\cal C}_2$ doesn't act on $\theta$(See also (\ref{calL}) or
(\ref{c2appen})), 
one multiplies the
expression (\ref{sol}) by any function of $\theta$ or $X^8$. 
Let us write down the eigenmodes as 
\bea
Y(y^m) = Y_{(0, \mu)}(X^A, A\neq 8) \, H(u), \qquad u \equiv
\cos^2 \theta.
\label{ans}
\eea
In the tensor product of ${\bf 7} \times {\bf 7} = [{\bf 1} \oplus
{\bf 27}]_S \oplus [{\bf 7} \oplus {\bf 14}]_A$ between the defining
representation ${\bf 7}=(0, 1)$ of $G_2$, the tensors transforming as
the symmetric part are given by \cite{Macfarlane}
\bea
Y_{{\bf 1}(0,0)AB}  & = & \frac{1}{7} \delta_{AB} \delta_{CD} V^C W^D,
\nonu \\
Y_{{\bf 27}(0,2)AB}  & = & \left[ \frac{1}{2} \left(\delta_{AC} 
\delta_{BD} +\delta_{AD}
    \delta_{BC} \right) - \frac{1}{7} \delta_{AB}
  \delta_{CD} \right] V^C W^D,
\label{tensors}
\eea
where $V^A$ and $W^B$ transform as ${\bf 7}$ of $G_2$ and have nothing
to do with $X^A$. Then one sees
that $Y_{{\bf 27}(0,2)AB}$ plays the role of $C_{AB}^{(0,2)}$ in (\ref{sol}) because 
$Y_{{\bf 27}(0,2)AB}$ is symmetric under the indices $A$ and $B$ and
traceless due to the fact that $\sum_{A=1}^7 Y_{{\bf 27}(0,2)AA} =0$
from (\ref{tensors}) \footnote{Although the full solution preserves
  the $G_2$ symmetry, the metric preserves a bigger $SO(7)$ symmetry
  from (\ref{7dmet3}). Since the KK modes see only the metric
  background, one can take the different approach by using the fact
  that the Laplacian eigenvalue problem on ${\bf S}^6$ is known. The
  Killing vector associated to the $SO(7)$ symmetry can be obtained
  from (\ref{Killing}) by taking
$21$ generators of $SO(7)$, that are $7\times 7$ matrices, as
  $(T_{ij})_{pq}=
\frac{1}{2}(\delta_{ip} \delta_{jq}-\delta_{iq} \delta_{jp})$ with the
property $\sum_{i,j=1}^7(T_{ij})^2=-\frac{3}{2}$. Then
one can check that the quadratic Casimir operator of $SO(7)$
corresponding to (\ref{c2}) is identical to the expressions of inside bracket 
in (\ref{c2appen}). Also  the eigenfunctions (\ref{sol}) can be
factorized 
by $\sin^{\mu} \theta \, Y_{\mu}$ where $Y_{\mu}$ is so-called $SO(7)$
spherical harmonics which depend on $(\theta_1, \alpha_1, \alpha_2,
\alpha_3, \theta_5, \theta_6)$. When we act the above quadratic
Casimir of $SO(7)$ on this spherical harmonics $Y_{\mu}$, we get 
$-\mu(\mu+5) Y_{\mu}$, as usual. We thank the referee for raising this
point.}.

Now let us solve the differential equation (\ref{diff1}) with (\ref{ans}). 
The differential operator (\ref{diffop}) can be computed explicitly
and it is given in the Appendix A (\ref{calL}) in terms of the angular variables.
Also this differential operator can be expressed in terms of the
rectangular coordinates via (\ref{c2appen}) and (\ref{remain}).
Since the quadratic Casimir operator has explicit eigenvalues of 
(\ref{eigenvalues}), it is obvious that the eigenvalue equation
(\ref{diff1})
can be solved more easily by using the rectangular coordinates rather
than the angular coordinates.   
When ${\cal C}_2$ acts on $Y(y^m)$ in (\ref{ans}), the only nonzero
contributions arise if it acts on $Y_{(0, \mu)}(X^A)$
because ${\cal C}_2$ doesn't act on $u$ as before. 

On the other hand, 
the remaining other piece (\ref{remain}) of ${\cal L}$ consisting of the second
derivative and first derivative with respect to 
$\theta$ or $u$ can act on either $ Y_{(0, \mu)}(X^A)$  
or $H(u)$. 
The former arises because the $X^A$($A=1, 2, \cdots,
7$) depends on the variable $\theta$ from (\ref{rect}). 
Since $ Y_{(0, \mu)}(X^A)$ is written in terms of rectangular
coordinates, it is convenient to rewrite the remaining operator in
terms of rectangular coordinates also using the chain rules, 
as in (\ref{remain}). One can compute the action of this remaining
operator on $ Y_{(0, \mu)}(X^A)$ explicitly. Moreover when 
the remaining operator acts on the function $H(u)$, then 
one obtains the second and first derivatives of $H(u)$ with respect to $u$. 
In this way, 
one performs these computations for $\mu =0, 1, 2, 3$ and expects the
final expression for general $\mu$. Actually we have checked this for
$\mu=4$. It turns out that 
the eigenvalue problem leads to the following differential equation 
\bea
(1-u) u H'' + \left[ c- (a_{+} + a_{-} +1) u \right] H' - a_{+} a_{-}
H =0,
\label{diff}
\eea
where the primes denote the derivatives with respect to $u$ and we
introduce the following quantities
\bea
a_{\pm} \equiv \frac{1}{6} \left( 9 + 3\mu \pm \sqrt{ 6 \mu^2 + 30 \mu
  + 81 +\frac{72}{5} m^2 \hat{L}^2 } \right), \qquad c=\frac{1}{2},
\qquad
\frac{2\sqrt{\frac{36\sqrt{2}\, 3^{\frac{1}{4}}}{25 \sqrt{5}}}}{L} 
\equiv \frac{1}{\hat{L}}.
\label{apm}
\eea
The complicated 
quantity $\sqrt{\frac{36\sqrt{2}\, 3^{\frac{1}{4}}}{25 \sqrt{5}}}$ is
the superpotential at the IR critical point and from the domain wall
equation, the scale factor $A(r)$ behaves as 
$A(r) \sim \frac{2\sqrt{\frac{36\sqrt{2}\, 3^{\frac{1}{4}}}{25
      \sqrt{5}}}}{L} r$ at the IR end of the flow \cite{Ahn0806n1} 
and we introduce a new
quantity $\hat{L}$ above in the spirit of \cite{KPR}. 
Note that the contribution from  ${\cal C}_2$ occurs only in the
term of $H$ in (\ref{diff}) while the contributions from the remaining
operator occur all the terms in $H'', H'$ or $H$ of (\ref{diff}). 
We have also checked that the eigenvalue equation (\ref{diff1}) holds 
(\ref{diff}) when ${\cal L}$ and $Y(y^m)$ are written 
in terms of angular coordinates via (\ref{calL}), (\ref{sol}),
(\ref{ans}) and (\ref{rect}): tracelessness of $(0,\mu)$ tensor is crucial.

The solution for (\ref{diff}) which is regular at $u=0$ is given by
\bea
H(u) = \HGF{a_-}{a_+}{c}{u},
\label{hyper}
\eea
which is convergent  for $|u| < 1$ for arbitrary $a_{-}, a_{+}$ and $c$.
When $a_{-} = -j$ for nonnegative integer $j$, this hypergeometric function
becomes a polynomial in $u$ of order $j$. Then the KK spectrum 
of minimally coupled scalar can be obtained by putting $a_{-}=-j$ in
(\ref{apm})
and solving for $m^2$. Then the mass-squared in $AdS_4$ can be
written, in terms of $\mu$ and $j$, as
\bea
m^2 = \frac{5}{24\hat{L}^2} \left( 12j^2 + 12 j \mu + 36j + \mu^2 + 8 \mu\right).
\label{square}
\eea
Plugging (\ref{hyper}) into (\ref{ans}) together with 
(\ref{apm}), one obtains the full eigenfunctions
\bea
Y(y^m) = C_{i_1 i_2 \cdots i_{\mu}}^{(0,\mu)} 
\left(\prod_{k=1}^{\mu} X^{i_{k}} \right) \HGF{-j}{3+\mu+j}{\frac{1}{2}}{(X^8)^2}.
\label{yym}
\eea
In this case, the hypergeometric functions are polynomials and let us
write down few cases according to the KK excitation number $j$
\bea
 j  &=&  0: \qquad Y(y^m) \sim Y_{(0, \mu)}(X^A), \nonu \\
 j  &=&  1:  \qquad Y(y^m) \sim  Y_{(0,
  \mu)}(X^A)\left[1-2(4+\mu)(X^8)^2 
\right], \nonu \\
 j  &=&  2: \qquad Y(y^m) \sim Y_{(0, \mu)}(X^A) 
\left[  1-4(5+\mu)(X^8)^2 + \frac{4}{3} (5+\mu)(6+\mu)(X^8)^4 \right],
\nonu \\
 j  &=&  3:  \qquad Y(y^m) \sim Y_{(0, \mu)}(X^A) \left[ 
1-6(6+\mu)(X^8)^2 +4(6+\mu)(7+\mu) (X^8)^4 \right. \nonu \\
& & \;\;\;\;\;\;\;\;\;\;\;\;\;\;\;\;\;\;\;\;\;\;\;\;\;\;\;-\left.
\frac{8}{15}(6+\mu)(7+\mu)(8+\mu)(X^8)^6\right],
\label{fewjs}
\eea
where $A\neq 8$.
The appearance of hypergeometric function in the eigenfunction for the
7-dimensional Laplacian operator is not so special and one sees the
similar feature in the compactification of $Q^{1,1,1}$ space \cite{Ahn99}. 

The dimension of the CFT operators dual to the KK modes (\ref{yym})
can be determined from the usual AdS/CFT correspondence \cite{Maldacena}
\bea
\Delta(\Delta-3) = m^2 \hat{L}^2.
\label{Del}
\eea
The $OSp(1|4)$ supermultiplets with spin-2 components  
are massless graviton multiplet($s_0=\frac{3}{2}$) with
$D(3,2)$ denoted by Class 3 of \cite{Heidenreich} 
and massive  graviton multiplet($s_0=\frac{3}{2}$) with
$D(E_0+\frac{1}{2},2)$ where $E_0 > \frac{5}{2}$ denoted by 
Class 4 of \cite{Heidenreich}.
As recognized in \cite{Ahn0806n1}, the massless graviton multiplet has
conformal dimension $\Delta=3$(the ground state component has
dimension $\Delta_0=\frac{5}{2}$ and see the Table 5 of \cite{Ahn0806n1}).
This ${\cal N}=1$ massless graviton multiplet characterized by 
$SD(\frac{5}{2},\frac{3}{2}|1)$
originates from the ${\cal N}=2$ massless graviton multiplet 
characterized by $SD(2,1,0|2)$ where the element zero stands for
$U(1)_R$ charge \cite{Gualtieri}. 
Similarly ${\cal N}=1$ massive graviton multiplet $SD(E_0
+\frac{1}{2}, \frac{3}{2}|1)$ with $E_0 \geq 2$ 
originates from 
the ${\cal N}=2$ massive long graviton multiplet or massive short
graviton multiplet.

Based on the findings (\ref{yym}), one can study 
the boundary operators dual to the KK modes by (\ref{yym}).
The theory has matter multiplet in seven flavors $\Phi^1, \Phi^2,
\cdots, \Phi^7$ transforming in the adjoint with flavor symmetry under
which the matter multiplet forms a ${\bf 7}$ of $G_2$ of the ${\cal N}=1$
theory \cite{Ahn0806n1}. 
The $\Phi^8$ is a singlet ${\bf 1}$ of $G_2$. The gauge theory 
conjectured to be dual to the $G_2$ ${\cal N}=1$ supergravity
background in this paper is a deformation of ABJM theory \cite{ABJM} by a
superpotential term quadratic in $\Phi^8$. The gauge theory also has
$G_2$ symmetry where $G_2$ symmetry corresponds to the global
rotations of $\Phi^1, \Phi^2, \cdots, \Phi^7$ into one another.
One identifies the $\Phi^A$ fields where $A\neq 8$ with the
coordinates $X^A$($A\neq 8$) and $\Phi^8$ with the coordinate $X^8$
up to normalization. One can read off the dual operators corresponding
to each of the KK modes. 

In Table 1 \footnote{In these three-dimensional SCFTs, there exist
monopole operators \cite{KKM} available for $k=1, 2$ that should be used to
obtain the correct gauge invariant operators. In the same spirit of
\cite{KPR}, we present only the ``schematic'' expressions of the dual
gauge theory operators which do not contain these monopole
operators. The corresponding ${\cal N}=1$ SCFT operators in
Chern-Simons matter theory hold for the
gauge group $U(2) \times U(2)$ with $k=1,2$. }, 
we present a few of these
modes and also provide the structure of the dual gauge theory
operators from (\ref{yym}). 
The branching rule for $SO(8)$ into $G_2$ is given by (\ref{sog2}).
The quantum number $\mu$ of $G_2$ is characterized by the Dynkin 
label $(0, \mu)$ as before. The KK excitation mode 
$j$ is nonnegative integer and this makes the hypergeometric funtion
be finite. The conformal dimension of dual SCFT operator is given by 
(\ref{Del}). Once the mass-squared formula is used via 
(\ref{square}), then this conformal dimension is fixed.
Starting with the ${\cal N}=1$ SCFT operator denoted by
$\Phi_{\alpha \beta \gamma}$ 
corresponding to the massless graviton multiplet, one can construct a
tower of KK modes by multiplying $\Phi^A(A\neq8)$ for $j=0$ modes.  
In general, one expects that for general quantum number $\mu$ of
$G_2$, the operator is given by the product of   
$\Phi_{\alpha \beta \gamma}$ with $\mu$ factors of $\Phi^A(A\neq8)$
where the $\mu$ indices are symmetrized.
For nonzero $j$'s, the explicit form (\ref{fewjs}) 
of hypergeometric functions is
useful to identify the corresponding ${\cal N}=1$ SCFT operators. 
For general $j$, there exists a 
polynomial up to the order $2j$ in $\Phi^8$ multiplied by 
$\Phi_{\alpha\beta\gamma}$.

\begin{table} 
\begin{center}
\begin{tabular}{|c|c|c|c|c|c|} \hline
$SO(8)$  & $G_2$
& $j$  & $\Delta$ & $m^2 \hat{L}^2$ & ${\cal N}=1$ \mbox{SCFT Operator}  \\
\hline
${\bf 1}(0,0,0,0)$ &   
${\bf 1}(0,0)$ & 0  & $3$  & 0 & $\Phi_{\alpha \beta \gamma}$ \\
\hline 
${\bf 8}_v(1,0,0,0)$& ${\bf 1}(0,0)$ & 1 &5 & 10 & $\Phi_{\alpha \beta
  \gamma} 
\left[1- 8(\Phi^8)^2 \right]$ \\
& ${\bf 7}(0,1)$ & 0  & $\frac{1}{4}(6+\sqrt{66}) $ & 
$\frac{15}{8}$ & $\Phi_{\alpha \beta \gamma} \Phi^A$ \\
\hline
$ {\bf 35}_v(2,0,0,0)$ 
&  ${\bf 1}(0,0)$ & $2$ & $\frac{1}{2}(3+\sqrt{109})$  & $25$ 
& $\Phi_{\alpha \beta \gamma} \left[ 1-20(\Phi^8)^2 +40(X^8)^4 \right]$ \\
& ${\bf 7}(0,1)$ & 1 & $\frac{1}{4}(6+\sqrt{266})$ 
& $\frac{115}{8}$ & $\Phi_{\alpha \beta \gamma} \Phi^A  
\left[1-10(\Phi^8)^2\right] $  \\
&  ${\bf 27}(0,2)$ & $0$ &$\frac{1}{6}(9+\sqrt{231}) $ 
& $\frac{25}{6}$ &  $\Phi_{\alpha \beta \gamma} \Phi^{(A} \Phi^{B)}$  \\
\hline
${\bf 112}_v(3,0,0,0)$ 
& ${\bf 1}(0,0)$ & $3$ & $\frac{3}{2}(1+\sqrt{21})$  
& $45$ & $\Phi_{\alpha \beta \gamma} \left[ 1-36(\Phi^8)^2 +168(\Phi^8)^4 -
\frac{896}{5} (\Phi^8)^6\right]$  \\
& ${\bf 7}(0,1)$ & $2$ & $\frac{1}{4}(6+\sqrt{546})$ 
 & $\frac{255}{8}$ & $\Phi_{\alpha \beta \gamma} \Phi^A \left[ 1-
   24(\Phi^8)^2 + 56(\Phi^8)^4\right]$ \\
& ${\bf 27}(0,2)$ & $1$ &$\frac{1}{6}(9+\sqrt{771})$ & $\frac{115}{6}$ 
&  $\Phi_{\alpha \beta \gamma} \Phi^{(A} \Phi^{B)} \left[
  1-12(\Phi^8)^2 \right]$  \\
&  ${\bf 77}'(0,3)$ & $0$ &$\frac{1}{4}(6+\sqrt{146})$ 
& $\frac{55}{8}$ & $\Phi_{\alpha \beta \gamma} \Phi^{(A} \Phi^B \Phi^{C)}$ \\
\hline
\end{tabular} 
\end{center}
\caption{\sl 
The first few spin-2 components of the massive(and massless) graviton multiplets. For each
multiplet we present $SO(8), G_2$ Dynkin labels (\ref{sog2}), the KK
excitation number $j$, the dimension $\Delta$ (\ref{Del}) of the spin-2 component
of the multiplet, the mass-squared $m^2 \hat{L}^2$ (\ref{square}) 
of the $AdS_4$ field
and the corresponding dual SCFT operator.}
\end{table} 

According to the observation of \cite{BHPW}, there exists a supersymmetric
RG flow from 
${\cal N}=1$ $G_2$-invariant fixed point to ${\cal N}=2$ $SU(3) \times
U(1)_R$-invariant fixed point. 
One describes the branching rules \cite{ps} of $G_2$ into $SU(3)$ as follows:
\bea
G_2 & \rightarrow & SU(3), \nonu \\
{\bf 1}(0,0) & \rightarrow& {\bf 1}(0,0), \nonu \\
{\bf 7}(0,1) & \rightarrow & {\bf 3}(1,0) \oplus \overline{{\bf 3}}(0,1)
\oplus {\bf 1}(0,0), \nonu \\
{\bf 27}(0,2) & \rightarrow & {\bf 6}(2,0) \oplus {\bf 8}(1,1) \oplus
\overline{{\bf 6}}(0,2)\oplus {\bf 3}(1,0) \oplus \overline{{\bf 3}}(0,1)
\oplus {\bf 1}(0,0), 
\nonu \\
{\bf 77}'(0,3) & \rightarrow & {\bf 10}(3,0) \oplus \overline{{\bf
    15}}(1,2)\oplus \overline{{\bf 10}}(0,3)\oplus {\bf 15}(2,1) \oplus
{\bf 6}(2,0) \oplus {\bf 8}(1,1)\oplus \overline{{\bf 6}}(0,2) \nonu \\
& & \oplus 
{\bf 3}(1,0)\oplus \overline{{\bf 3}}(0,1)\oplus {\bf 1}(0,0),
\nonu \\
(0,\mu) & 
\rightarrow & [(\mu,0) \oplus (\mu-1, 1) \oplus \cdots (0,\mu)] \oplus
[(\mu-1,0) \oplus (\mu-2, 1) \oplus \cdots (0,\mu-1)] 
\nonu \\
&& \oplus \cdots \oplus [(1,0) \oplus (0,1)] \oplus (0,0),
\label{g2su3}
\eea
where the dimension \cite{Macfarlane} of $G_2$ 
representation $(0,\mu)$ is given by $\frac{1}{120} 
\prod_{i=1}^{4}({\bf \mu +i})({\bf 2\mu+5})$.
At the level of $\mu=0$ excitation, this is the familiar massless
graviton multiplet in $AdS_4$ and corresponds to the stress energy
tensor in the SCFT. In the standard ${\cal N}=2$ stress energy
superfield ${\cal T}_{\alpha \beta}^{(0)}$ in the notation of \cite{KPR},
the components are given by a vector boson that is related to 
$U(1)_R$ symmetry, two fermionic supersymmetry
generators, and energy momentum tensor.  
Then we denote $D_{\alpha} {\cal T}_{\beta \gamma}^{(0)}$
corresponding to the ${\cal N}=1$ massless graviton multiplet
by $\Phi_{\alpha \beta \gamma}$ which has ${\cal N}=1$ supercurrent
and the energy momentum tensor in components \cite{Ahn0806n1}. 
Here $D_{\alpha}$ is an
${\cal N}=1$ superderivative. 
As explained before, the massless graviton multiplet($s_0=\frac{3}{2}$
in ${\cal N}=1$ notation) with
$D(3,2)$ is classified  by the Class 3 of \cite{Heidenreich} 
and has
conformal dimension $\Delta=3$(the ground state component has
dimension $\Delta_0=\frac{5}{2}$).

Now let us describe the massive graviton multiplet.
At the level of $\mu=1$ excitation, the ${\bf 7}$ representation of
$G_2$ breaks into three different representations of $SU(3)$ as
above. 
In ${\cal N}=2$
theory, ${\bf 3}$ representation of $SU(3)$ corresponds to ${\cal Z}^A$ where
$A=1,2,3$
and its complex conjugates $\overline{\bf 3}$ representation 
of $SU(3)$ corresponds to $\overline{\cal Z}_A$ 
\cite{Ahn0806n2,BKKS,KKM,KPR}. 
Then the ${\cal N}=1$ SCFT
operator goes to the ${\cal N}=2$ 
SCFT operators \cite{KPR} schematically as follows:
\bea
\Phi_{\alpha \beta \gamma} \Phi^A \rightarrow 
{\cal T}_{\alpha
\beta}^{(0)} \left[ {\cal Z}^A \oplus \overline{\cal Z}_A \oplus  
( {\cal Z}^4 + \overline{\cal Z}_4) \right],
\label{7}
\eea
where ${\cal Z}^4$ and
$\overline{\cal Z}_4$ are 
singlets of $SU(3)$ and the sum of these corresponds to the ${\cal
N}=1$ superfield $\Phi^7$. 
According to the observation of \cite{KPR}, the last representation of
(\ref{7})
belongs to the ${\cal N}=2$ short graviton multiplet.
The other representations of (\ref{7}) 
belong to the ${\cal N}=2$ long graviton 
multiplet.
For $j=1$ case in the singlet ${\bf 1}$ of $G_2$, 
the structure of ${\cal N}=1$
SCFT operator can be obtained from the expression of hypergeometric
function for $j=1$ case (\ref{fewjs}). 

At the level of $\mu=2$ excitation, the ${\bf 27}$ representation of
$G_2$ breaks into six different representations of $SU(3)$.  
Then the ${\cal N}=1$ SCFT
operator goes to the ${\cal N}=2$ 
SCFT operators \cite{KPR} with the same order for $SU(3)$
representations 
(\ref{g2su3})
\bea
\Phi_{\alpha \beta \gamma} \Phi^{(A} \Phi^{B)}  \rightarrow & &
{\cal T}_{\alpha
\beta}^{(0)} \left[ {\cal Z}^{(A} {\cal Z}^{B)} \oplus 
( {\cal Z}^A \overline{\cal Z}_B -\frac{1}{3}
\delta_{B}^{A} {\cal Z}^C \overline{\cal Z}_C) \oplus
\overline{\cal Z}_{(A}  \overline{\cal Z}_{B)} \right. \nonu \\
& & \oplus  \left.
{\cal Z}^A({\cal Z}^4 + \overline{\cal Z}_4)
\oplus  \overline{\cal Z}_A({\cal Z}^4 + \overline{\cal
  Z}_4) \oplus  ({\cal Z}^4 + \overline{\cal Z}_4)^2 \right].
\label{27}
\eea
The right hand side represents each $SU(3)$ representation, term by term, in 
(\ref{g2su3}) exactly. The second representation can be obtained from
the tensor product of ${\bf 3}$ and $\overline{\bf 3}$ which leads to
${\bf 8}$ and corresponds to the massless vector multiplet. 
On the other hand, the ${\cal N}=1$ massless vector multiplet has
conformal dimension $\Delta=2$ for spin-1(the ground state component has
dimension $\Delta_0=\frac{3}{2}$ and see the Table 5 of \cite{Ahn0806n1}).
This ${\cal N}=1$ massless vector multiplet characterized by 
$SD(\frac{3}{2},\frac{1}{2}|1)$
originates from the ${\cal N}=2$ massless vector multiplet 
characterized by $SD(1,0,0|2)$ \cite{Gualtieri}. 
The last three representations of (\ref{27}) 
are obtained from (\ref{7}) by multiplying 
$({\cal Z}^4 +\overline{\cal Z}_4)$.
The last representation of
(\ref{27})
belongs to the ${\cal N}=2$ short graviton multiplet \cite{KPR}.
The first three representations can be obtained from the fact that the
product $\Phi^{(A} \Phi^{B)}$ has three possible cases: two products of
${\cal Z}^A$'s, two products $\overline{\cal Z}_A$'s and the product
of ${\cal Z}^A$ and $\overline{\cal Z}_A$.
For $j=1, 2$ cases in the ${\bf 7, 1}$ of $G_2$, 
the structure of ${\cal N}=1$
SCFT operator is read off from the expression of hypergeometric
function for $j=1, 2$ cases (\ref{fewjs}). 

Finally, at the level of $\mu=3$ excitation, the ${\bf 77}'$ representation of
$G_2$ breaks into ten different representations of $SU(3)$.  
Then the ${\cal N}=1$ SCFT
operator
goes to the ${\cal N}=2$ 
SCFT operators \cite{KPR}
\bea
\Phi_{\alpha \beta \gamma} \Phi^{(A} \Phi^{B} \Phi^{C)}
 \rightarrow & & 
{\cal T}_{\alpha
\beta}^{(0)} \left[ {\cal Z}^{(A} {\cal Z}^{B} {\cal Z}^{C)} \oplus
( \overline{\cal Z}_{(A} \overline{\cal Z}_{B)} {\cal Z}^C
  -\frac{1}{3}
\delta^C_{(A} \overline{\cal Z}_{B)} \overline{\cal Z}_D {\cal Z}^D) \oplus
\overline{\cal Z}_{(A}  \overline{\cal Z}_{B}
\overline{\cal Z}_{C)} \right. \nonu \\
&& \oplus  
( {\cal Z}^{(A} {\cal Z}^{B)} \overline{\cal Z}_C
  -\frac{1}{3}
\delta_C^{(A} {\cal Z}^{B)} {\cal Z}^D \overline{\cal Z}_D) \oplus
{\cal Z}^{(A} {\cal Z}^{B)} 
( {\cal Z}^4 + \overline{\cal Z}_4) \nonu \\
&& \oplus     ( {\cal Z}^A \overline{\cal Z}_B -\frac{1}{3}
\delta_{B}^{A} {\cal Z}^C \overline{\cal Z}_C)
( {\cal Z}^4 + \overline{\cal Z}_4) \oplus
\overline{\cal Z}_{(A} \overline{\cal Z}_{B)} 
( {\cal Z}^4 + \overline{\cal Z}_4) \nonu \\
&& \oplus  \left. 
{\cal Z}^A ({\cal Z}^4 + \overline{\cal Z}_4)^2
\oplus  \overline{\cal Z}_A ({\cal Z}^4 + \overline{\cal
  Z}_4)^2 \oplus  ({\cal Z}^4 + \overline{\cal Z}_4)^3
 \right].
\label{77}
\eea
The second representation in the right hand side can be obtained from
the tensor product of ${\bf 8}$ and $\overline{\bf 3}$ which leads to
$\overline{\bf 15}$.
The first four representations can be obtained from the fact that the
product $\Phi^{(A} \Phi^{B} \Phi^{C)}$ has four possible cases: three products of
${\cal Z}^A$'s, three products of $\overline{\cal Z}_A$'s, the product
of two ${\cal Z}^A$'s and $\overline{\cal Z}_A$ and 
the product of ${\cal Z}^A$ and two $\overline{\cal Z}_A$'s.
The last six representations of (\ref{77}) 
are obtained from (\ref{27}) by multiplying 
$({\cal Z}^4 +\overline{\cal Z}_4)$.
The last representation of
(\ref{77})
belongs to the ${\cal N}=2$ short graviton multiplet \cite{KPR}
and 
the other representations of (\ref{77}) 
belong to the ${\cal N}=2$ long graviton 
multiplet.
For $j=1, 2, 3$ cases in ${\bf 27, 7, 1}$ of $G_2$, 
the structure of ${\cal N}=1$
SCFT operator is read off from the expression of hypergeometric
function for $j=1, 2, 3$ cases (\ref{fewjs}). 

\section{
Conclusions and outlook }

We computed the KK reduction for spin-2 excitations around the warped
11-dimensional theory background that is dual to the ${\cal N}=1$
mass-deformed Chern-Simons matter theory with $G_2$ symmetry. 
The spectrum of spin-2 excitations was given by solving the equations
of motion for minimally coupled scalar theory in this background.
The $AdS_4$ mass formula of the KK modes is given by (\ref{square})
and the corresponding wavefunctions on the 7-dimensional manifold are
given by (\ref{yym}). The quantum number $\mu$ for $G_2$
representation and the KK excitation number $j$ arise in this mass formula.      
We calculated the dimensions of the dual operators in the boundary
SCFT via AdS/CFT correspondence and in Table 1 we presented the
summary of this work.

In the classification of \cite{Heidenreich},
the massive multiplets for lower spins arise also. For example,
the $OSp(1|4)$ supermultiplets with  spin-$\frac{3}{2}$ components
are massless gravitino multiplet($s_0=1$) with
$D(\frac{5}{2},\frac{3}{2})$ denoted by Class 3
and massive  gravitino multiplet($s_0=1$) with
$D(E_0+\frac{1}{2},\frac{3}{2})$ where $E_0 > 2$ denoted by 
Class 4. 
The ${\cal N}=1$ massive gravitino multiplet $SD(E_0, 1|1)$ with $E_0 \geq 2$ 
originates from 
the ${\cal N}=2$ massive long gravitino multiplet.
The massive ${\cal N}=8$ supermultiplet \cite{DNP} at level $n$ for
spin-$\frac{3}{2}$
is given by $SO(8)$ Dynkin labels $(n, 0, 0, 1) \oplus (n-1, 0, 1, 0)$. 
Definitely this provides the gauge theory operators dual to
lower spin excitations. However, one should find out the right form for the 
perturbations that decouple from all other perturbations.

\vspace{.7cm}

\centerline{\bf Acknowledgments}

We would like to thank 
Sangmin Lee and Fabio D. Rocha 
for discussions. 

\appendix

\renewcommand{\thesection}{\large \bf \mbox{Appendix~}\Alph{section}}
\renewcommand{\theequation}{\Alph{section}\mbox{.}\arabic{equation}}
\section{The differential operator, quadratic Casimir operator of $G_2$, and
  the generators of $G_2$ }

The differential operator acting on 7-dimensional ellipsoid 
is given by (\ref{diffop}) and this can be written, from the metric
(\ref{7dmetric}) and the warp factor (\ref{delta}) with (\ref{ab}) and
(\ref{xi}),  in terms of angular
coordinates
as follows:
\bea
\left(\frac{2}{5} \right)^{-3/4} 3^{-5/8} {\cal L} & = & 
\frac{18}{5}  \pa_{\theta}^2 +
 \frac{108}{5} c_{\theta} s^{-1}_{\theta} \pa_{\theta}
\nonu \\
&+&    \frac{6}{5} \left(2
  +c_{2\theta} 
\right) s^{-2}_{\theta}  \left[ s^{-2}_{\theta_6} \pa^2_{\theta_1} +
4    s^{-2}_{\theta_1} s^{-2}_{\theta_6} \pa^2_{\alpha_1}
+ 4     s^{-2}_{\alpha_1}  
s^{-2}_{\theta_1} s^{-2}_{\theta_6} \pa^2_{\alpha_2} +
30 s^{-1}_{\theta_6} c_{\theta_6} 
\pa_{\theta_6} \right.
\nonu \\
&- & 8    c_{\alpha_1} s^{-2}_{\alpha_1} 
 s^{-2}_{\theta_1} s^{-2}_{\theta_6} \pa_{\alpha_2} \pa_{\alpha_3}
-
4     s^{-2}_{\alpha_1} \left(
  c^2_{\alpha_1}-s^{-2}_{\theta_1} 
\right) c^{-2}_{\theta_1} s^{-2}_{\theta_6} \pa^2_{\alpha_3}
-4     
c^{-2}_{\theta_1} s^{-2}_{\theta_6} \pa_{\alpha_3} \pa_{\theta_5}
\nonu \\
&+ &    \left.
c^{-2}_{\theta_1} s^{-2}_{\theta_6} \pa^2_{\theta_5}
+    \pa^2_{\theta_6}
+
 \left( 1+ 2 c_{2\theta_1}\right)
 s^{-1}_{\theta_1} s^{-2}_{\theta_6} s^{-1}_{\theta_1}
\pa_{\theta_1}
+
4 
 s^{-1}_{\alpha_1} c_{\alpha_1} s^{-2}_{\theta_6} s^{-2}_{\theta_1}
\pa_{\alpha_1}\right]
\nonu \\
&=&
\frac{18}{5}  \pa_{\theta}^2 +
 \frac{108}{5} c^{-1}_{\theta} s^{-1}_{\theta} \pa_{\theta} +
 \frac{9}{40}  \left(2
  +c_{2\theta} 
\right) s^{-2}_{\theta} {\cal C}_2,
\label{calL}
\eea
where the quadratic Casimir operator (\ref{c2}) with (\ref{Killing}) can be
written as 
\bea
{\cal C}_2  & = & \frac{16}{3}   \left( \sum_{i, j=1, i\neq j}^{7} 
(X^i)^2  \pa^2_{X^j} - 2 \sum_{i, j=1, j > i}^{7} X^i X^j \pa_{X^i} \pa_{X^j}
-6 \sum_{i=1}^{7} X^i \pa_{X^i} \right).
\label{c2appen}
\eea
So it is obvious that the action of this ${\cal C}_2$ on the function
$H(u)$ vanishes because the right hand side of (\ref{c2appen}) doesn't
contain any differential operator on the variable $X^8$.
The remaining parts of (\ref{calL}) can be written in terms of
rectangular coordinates as follows:
\bea
\pa_{\theta}^2 +
 6 c^{-1}_{\theta} s^{-1}_{\theta} \pa_{\theta} & = &
c^2_{\theta} s^{-2}_{\theta} \sum_{i=1}^{7} (X^i)^2 \pa^2_{X^i} 
-\sum_{i=1}^{7} X^i X^8 \pa_{X^i} \pa_{X^8} + s^2_{\theta} \pa^2_{X^8}
\nonu \\
& + & \left(-1+7c^2_{\theta}\right) s^{-2}_{\theta} \sum_{i=1}^{7} X^i
\pa_{X^i}-7s_{\theta} c_{\theta} \pa_{X^8}.   
\label{remain}
\eea
The first and fourth terms of (\ref{remain}) contribute to the terms
in linear of $H$, the second and last terms contribute to the terms 
of $H'$ and the third term contributes to the terms of $H''$ in (\ref{diff}).

The generators of $G_2$ can be chosen as real $7\times 7$
matrices. The explicit realization of the generators was presented in \cite{CCDOS}.
The embedding of $G_2$ in the group $SO(7)$ is generated by the $14$
elements $T^a, a=1,2, \cdots, 14$:
\bea
&& T^{1}=\left(\begin{array}{ccccccc}
0 & 0 &0  & 0 & 0 & 0 & 0\\{}
0 & 0 & 0 & 0 & 0 & 0 & 0\\{}
0 & 0 & 0 & 0 & 0 & 0 & 0\\{}
0 & 0 & 0 & 0 & 0 & 0 &-1 \\{}
0 & 0 & 0 & 0 & 0 & -1& 0 \\{}
0 & 0 & 0 & 0 & 1 & 0 & 0 \\{}
0 & 0 & 0 & 1 & 0 & 0 & 0
\end{array}\right),
\qquad
T^{2}=\left(\begin{array}{ccccccc}
0 & 0 &0  &0  & 0 &0 & 0\\{}
0 & 0 & 0 & 0 & 0 &0 & 0\\{}
0 & 0 & 0 & 0 & 0 & 0& 0\\{}
0 & 0 & 0 & 0 & 0 &1 &0 \\{}
0 & 0 & 0 & 0 &0  &0 &-1 \\{}
0 & 0 & 0 & -1& 0 &0 &0 \\{}
0 & 0 & 0 & 0 & 1 & 0& 0
\end{array}\right),
\nonu \\
&& T^{3}=\left(\begin{array}{ccccccc}
0 & 0 &0  &0  & 0  &0  & 0\\{}
0 & 0 & 0 & 0 & 0  &0  & 0\\{}
0 & 0 & 0 & 0 & 0  & 0 & 0\\{}
0 & 0 & 0 & 0 & -1 &0  &0 \\{}
0 & 0 & 0 & 1 &0   &0  &0 \\{}
0 & 0 & 0 & 0 & 0  &0  &-1 \\{}
0 & 0 & 0 & 0 & 0  & 1 & 0
\end{array}\right),
\qquad
T^{4}=\left(\begin{array}{ccccccc}
0 &0  &0  &0  & 0 &0 & 0\\{}
0 & 0 & 0 & 0 & 0 &0 & 1\\{}
0 & 0 & 0 & 0 & 0 & 1& 0\\{}
0 & 0 & 0 & 0 & 0 &0 &0 \\{}
0 & 0 & 0 & 0 &0  &0 &0 \\{}
0 & 0 & -1 & 0 & 0 &0 &0 \\{}
0 & -1 & 0 & 0 & 0 & 0& 0
\end{array}\right),
\nonu \\
&& T^{5}=\left(\begin{array}{ccccccc}
0 &0  &0  &0  & 0 &0 & 0\\{}
0 & 0 & 0 & 0 & 0 &-1 & 0\\{}
0 & 0 & 0 & 0 & 0 & 0& 1\\{}
0 & 0 & 0 & 0 & 0 &0 &0 \\{}
0 & 0 & 0 & 0 &0  &0 &0 \\{}
0 & 1 & 0 & 0 & 0 &0 &0 \\{}
0 & 0 & -1 & 0 & 0 & 0& 0
\end{array}\right),
\qquad
T^{6}=\left(\begin{array}{ccccccc}
0 &0  &0  &0  & 0 &0 & 0\\{}
0 & 0 & 0 & 0 & 1 &0 & 0\\{}
0 & 0 & 0 & -1 & 0 & 0& 0\\{}
0 & 0 & 1 & 0 & 0 &0 &0 \\{}
0 & -1 & 0 & 0 &0  &0 &0 \\{}
0 & 0 & 0 & 0 & 0 &0 &0 \\{}
0 & 0 & 0 & 0 & 0 & 0& 0
\end{array}\right),
\nonu \\
&& T^{7}=\left(\begin{array}{ccccccc}
0 &0  &0  &0  & 0 &0 & 0\\{}
0 & 0 & 0 & -1 & 0 &0 & 0\\{}
0 & 0 & 0 & 0 & -1 & 0& 0\\{}
0 & 1 & 0 & 0 & 0 &0 &0 \\{}
0 & 0 & 1 & 0 &0  &0 &0 \\{}
0 & 0 & 0 & 0 & 0 &0 &0 \\{}
0 & 0 & 0 & 0 & 0 & 0& 0
\end{array}\right),
\label{gen}
\eea
and
\bea
&&
T^{8}=\frac{1}{\sqrt{3}} \left(\begin{array}{ccccccc}
0 &0  &0  &0  & 0 &0 & 0\\{}
0 & 0 & -2 & 0 & 0 &0 & 0\\{}
0 & 2 & 0 & 0 & 0 & 0& 0\\{}
0 & 0 & 0 & 0 & 1 &0 &0 \\{}
0 & 0 & 0 & -1 &0  &0 &0 \\{}
0 & 0 & 0 & 0 & 0 &0 &-1 \\{}
0 & 0 & 0 & 0 & 0 & 1& 0
\end{array}\right),
\label{gen1} 
\\
&& T^{9}=\frac{1}{\sqrt{3}} \left(\begin{array}{ccccccc}
0 &-2  &0  &0  & 0 &0 & 0\\{}
2 & 0 & 0 & 0 & 0 &0 & 0\\{}
0 & 0 & 0 & 0 & 0 & 0& 0\\{}
0 & 0 & 0 & 0 & 0 &0 &1 \\{}
0 & 0 & 0 & 0 &0  &-1 &0 \\{}
0 & 0 & 0 & 0 & 1 &0 &0 \\{}
0 & 0 & 0 & -1 & 0 & 0& 0
\end{array}\right),
\qquad
T^{10}=\frac{1}{\sqrt{3}} \left(\begin{array}{ccccccc}
0 &-2  &0  &0  & 0 &0 & 0\\{}
0 & 0 & 0 & 0 & 0 &0 & 0\\{}
2 & 0 & 0 & 0 & 0 & 0& 0\\{}
0 & 0 & 0 & 0 & 0 &-1 &0 \\{}
0 & 0 & 0 & 0 &0  &0 &-1 \\{}
0 & 0 & 0 & 1 & 0 &0 &0 \\{}
0 & 0 & 0 & 0 & 1 & 0& 0
\end{array}\right),
\nonu \\
&& T^{11}=\frac{1}{\sqrt{3}} \left(\begin{array}{ccccccc}
0 &0  &0  &-2  & 0 &0 & 0\\{}
0 & 0 & 0 & 0 & 0 &0 & -1\\{}
0 & 0 & 0 & 0 & 0 & 1& 0\\{}
2 & 0 & 0 & 0 & 0 &0 &0 \\{}
0 & 0 & 0 & 0 &0  &0 &0 \\{}
0 & 0 & -1 & 0 & 0 &0 &0 \\{}
0 & 1 & 0 & 0 & 0 & 0& 0
\end{array}\right),
\qquad
T^{12}=\frac{1}{\sqrt{3}} \left(\begin{array}{ccccccc}
0 &0  &0  &0  & -2 &0 & 0\\{}
0 & 0 & 0 & 0 & 0 &1 & 0\\{}
0 & 0 & 0 & 0 & 0 & 0& 1\\{}
0 & 0 & 0 & 0 & 0 &0 &0 \\{}
2 & 0 & 0 & 0 &0  &0 &0 \\{}
0 & -1 & 0 & 0 & 0 &0 &0 \\{}
0 & 0 & -1 & 0 & 0 & 0& 0
\end{array}\right),
\nonu \\
&& T^{13}=\frac{1}{\sqrt{3}} \left(\begin{array}{ccccccc}
0 & 0 & 0 & 0 & 0  &-2 & 0\\{}
0 & 0 & 0 & 0 & -1 & 0 & 0\\{}
0 & 0 & 0 & -1 & 0 & 0 & 0\\{}
0 & 0 & 1 & 0 & 0  &0   &0 \\{}
0 & 1 & 0 & 0 &0   &0   &0 \\{}
2 & 0 & 0 & 0 & 0  &0   &0 \\{}
0 & 0 & 0 & 0 & 0  & 0  & 0
\end{array}\right),
\qquad
T^{14}=\frac{1}{\sqrt{3}} \left(\begin{array}{ccccccc}
0 &0  &0  &0  & 0 &0 & -2\\{}
0 & 0 & 0 & 1 & 0 &0 & 0\\{}
0 & 0 & 0 & 0 & -1 & 0& 0\\{}
0 & -1 & 0 & 0 & 0 &0 &0 \\{}
0 & 0 & 1 & 0 &0  &0 &0 \\{}
0 & 0 & 0 & 0 & 0 &0 &0 \\{}
2 & 0 & 0 & 0 & 0 & 0& 0
\end{array}\right),
\nonu
\eea
where the two matrices $T^3$ and $T^8$ generate the Cartan subgroup of
$G_2$.
There exist six $SU(2)$ subgroups generated by the elements 
$(T^1, T^2,T^3), (T^4,T^5,\frac{1}{2}(\sqrt{3} T^8 + T^3)), (T^6, T^7,
\frac{1}{2}(-\sqrt{3} T^8 + T^3)), (\sqrt{3} T^9,
\sqrt{3}T^{10},\sqrt{3} T^8), (\sqrt{3} T^{11},
\sqrt{3}T^{12},\frac{1}{2}(-\sqrt{3} T^8 + 3T^3))$ and $(\sqrt{3} T^{13},
\sqrt{3}T^{14},\frac{1}{2}(\sqrt{3} T^8 + 3T^3))$ from the fundamental
commutation relations between these generators \cite{GLORT}.


\end{document}